\documentstyle[12pt]{article}
\newcommand{\newc}{\newcommand} 
\newc{\ra}{\rightarrow} 
\newc{\lra}{\leftrightarrow} 
\newc{\beq}{\begin{equation}} 
\newc{\eeq}{\end{equation}} 
\newc{\barr}{\begin{eqnarray}} 
\newc{\earr}{\end{eqnarray}} 

 1

\begin{document} 
\begin{titlepage}

\vspace*{0.3cm}
\begin{center}
{\large \bf COLD DARK MATTER DETECTION VIA THE \\ LSP-NUCLEUS ELASTIC SCATTERING
\footnote{ Presented by J.D. Vergados}}

\vspace{0.8cm}
{J.D. VERGADOS and T.S. KOSMAS }

\vspace{0.4cm}
{\it Theoretical Physics Section, University of Ioannina, GR 451 10, Greece}
\end{center}

\vspace{0.4cm}
\abstract{The momentum transfer dependence of the LSP-nucleus elastic 
scattering cross sections is studied. New imput SUSY parameters obtained 
in a phenomenologically allowed parameter space are used to calculate
the coherent rate for various nuclear systems and the spin matrix elements 
for the proposed $^{207}Pb$ target. The results are compared to those 
obtained from other cold dark matter detection targets.}

\end{titlepage}


\section{Introduction }
Recently, the possibility to directly detect the lightest supersymmetric 
particle (LSP), a candidate for cold dark matter, via the recoiling of 
a nucleus in the elastic scattering process:
\beq
\chi \, +\, (A,Z) \, \ra \, \chi \,  + \, (A,Z)^* 
\label{eq:1.1}  
\eeq
($\chi$ denotes the LSP) has been proposed.~\cite{JDV,KVprd} In the present 
study we proceed as follows:

1) We write down the effective Lagrangian at the elementary particle 
(quark) level obtained in the framework of supersymmetry as described 
in refs.~\cite{JDV,KVprd}

2) We go from the quark to the nucleon level using an appropriate quark 
model for the nucleon. Special attention in this step is paid to the 
scalar couplings, which dominate the coherent part of the cross section
and the isoscalar axial currents which depend on the assumed quark model.

3) We compute the relevant nuclear matrix elements using 
as reliable as possible many boby nuclear wave function hopping
that, by putting as accurate nuclear physics input as possible, 
we will be able to constrain the SUSY parameters as much as possible.

4) We calculate the modulation of the cross sections due to the earth's
revolution around the sun and the motion of the sun around the center of
the galactic disc by a folding procedure.

Our goal is to calculate LSP-nucleus event rates for $Pb$ nucleus by using 
new imput SUSY parameters~\cite{Kane,Casta} obtained in a phenomenologically 
allowed parameter space.
We focus on the spin matrix elements of $^{207}Pb$, since this target, 
in addition to its experimental qualifications, has the advantage of a 
rather simple nuclear structure. Furthermore, its spin matrix element, 
especially the isoscalar one, does not exhibit large quenching as that 
of the light and up to now much studied $^{29}Si$ and $^{73}Ge$ 
nuclei.~\cite{Ress} We compare our results to those given from other 
cold dark matter detection targets.~\cite{Ress} 

The total cross section of the LSP-nucleus reaction (1) can be written 
as~\cite{KVprd}

\barr
\sigma &=& \sigma_0 (\frac{m_1}{m_p})^2 \frac{1}{(1+\eta)^2} \,
 \{ A^2 \, [[\beta^2 (f^0_V - f^1_V \frac{A-2 Z}{A})^2 
\nonumber \\ & + & 
(f^0_S - f^1_S \frac{A-2 Z}{A})^2 \, ]I_0(u) -
\frac{\beta^2}{2} \frac{2\eta +1}{(1+\eta)^2}
(f^0_V - f^1_V \frac{A-2 Z}{A})^2 I_1 (u) ]
\nonumber \\ & + & 
(f^0_A \Omega_0(0))^2 I_{00}(u) + 2f^0_A f^1_A \Omega_0(0) \Omega_1(0)
I_{01}(u) + (f^1_A \Omega_1(0))^2 I_{11}(u) \, \} 
\label{eq:2.1}
 \earr

\noindent 
where $m_1, \, m_p\,$ is the LSP, proton masses, $\eta = m_1/m_p A$ and
$\sigma_0 \simeq 0.77 \times 10^{-38}cm^2$. The parameters $f^{\tau}_S, 
\,\,f^{\tau}_V, \,\, f^{\tau}_A$, with $\tau =0,1$ an isospin index, 
describe the scalar, vector and axial vector couplings and are determined 
in various SUSY models~\cite{Kane,Casta} (see ref.~\cite{KVprd}). 
The momentum transfer enters via u as

\beq
u = \frac{1}{2} \left( \frac{2\beta m_1 c^2}{(1+\eta)} 
\frac{b}{\hbar c}\right)^2,
\qquad \beta = \frac{v}{c} \, \approx \, 10^{-3}
\label{eq:2.2}  
\eeq

\noindent 
(b is the harmonic oscillator parameter).
In Eq. (\ref{eq:2.1}), the terms in the square brackets describe the coherent
cross section and involve the integrals $I_{\rho}(u)$ ($\rho =0,1$)
while the other terms describe the spin dependence of the  
cross section by means of the integrals  $I_{\rho\rho^{\prime}}(u)$
($\rho,\rho^{\prime} =0,1$, isospin indices) which are normalized so 
as $I_{\rho}(0)=1$ and $I_{\rho \rho^{\prime}}(0)=1$. 
For their definitions see ref.~\cite{KVprd} where they are calculated 
by using realistic nuclear form factors.

In Fig. 1(a) we show the variation of $I_0(u)$ (the others behave similarly).
We see that $u$ can be quite big for large mass of the LSP
and heavy nuclei even though 
the energy transfer is small ($\le 100 KeV$). The total cross section can 
in such instances be reduced by a factor of about five. 

The spin matrix element of heavy nuclei like $^{207}Pb$ has not been 
previously evaluated, since one expects the relative 
importance of the spin versus the coherent mode to be more pronounced
on light nuclei. However, the spin matrix element in the 
light isotopes is quenched, while that 
of $^{207}Pb$ does not show large quenching. For this feature,
we recently proposed $^{207}Pb$ nucleus as an important candidate
in the LSP detection.~\cite{KVprd} Furthermore, this nucleus has some 
additional advantages
as: i) it is believed to have simple structure (one $2p 1/2$ neutron
hole outside the doubly magic nucleus $^{208}Pb$) and 
ii) it has low angular momentum and therefore only two multipoles $\lambda =0$
and $\lambda =2$ with a $J$-rank of $\kappa=1$ can contribute even at large
momentum transfers. 

The dependence of the cross section on the momentum transfer for 
$^{207}Pb$ has been extensively studied in ref.~\cite{KVprd} In Fig. 1(a) 
the variation of the integral $I_{11}(u)$, which describe the spin part 
of $^{207}Pb$, is shown.

\begin{center}
\begin{figure}

\vspace{6.0cm}

\vspace{0.3cm}

\caption{(a) Plot of the integrals $I_{11}(u)$ and $I_0(u)$ for $Pb$, 
where $u$ is given by Eq. (3). We see that $I_{11}$ is quite a bit less 
retarded compared to $I_{0}$. (b) $K^l_0$ integrals (for l=0 
and l=1) entering the dominant scalar part of the event rate.}
\end{figure}
\end{center}

\section{Folding of the cross section with a Maxwellian distribution}
Due to the revolution of the earth around the sun, the event rate
becomes modulated. This effect is studied by folding the total 
cross section Eq. (\ref{eq:2.1}) with an appropriate distribution. 
If we assume a Maxwell type distribution, which is consistent with
the velocity distribution of LSP into the galactic halo,
the counting rate for a target with mass $m$ takes the form~\cite{KVprd}

\beq
\Big<\frac{dN}{dt}\Big> =\frac{\rho (0)}{m_1} \frac{m}{Am_p} \sqrt{<v^2>}
<\Sigma>
\label{eq:3.1}  
\eeq

\noindent
where $\rho (0) = 0.3 GeV/cm^3$, the LSP density in our vicinity and
$<\Sigma>$ is given by

\barr
<\Sigma>&=&\Big(\frac{m_1}{m_p}\Big)^2 \frac{\sigma_0}{(1+\eta)^2}
\Big\{A^2 \Big[ <\beta^2> \\
\nonumber
&& \times \Big(f^0_V-f^1_V \frac{A-2 Z}{A})^2 
\Big(J_0-\frac{2\eta+1}{2(1+\eta)^2}J_1\Big) +
(f^0_S-f^1_S\frac{A-2 Z}{A})^2{\tilde J}_0\Big]  \\
\nonumber
&& + \Big( f^0_A \Omega_0(0)\Big)^2 J_{00}
+ 2 f^0_A f^1_A \Omega_0(0)\Omega_1(0) J_{01}
+ \Big( f^1_A \Omega_1(0)\Big)^2 J_{11} \Big\}
\label{eq:3.2}  
\earr

\noindent
The parameters ${\tilde J}_0$, $J_\rho$, $J_{\rho\sigma}$ describe the
scalar, vector and spin part of the velocity averaged counting rate, 
respectively. They are functions of $\beta_0 =v_0/c$, $\delta =2v_1/v_0$
and $u_0$, where $v_0$ is the velocity of the sun around the galaxy,
$v_1$ the velocity of the Earth around the sun and $u_0$ is given
by an expression like that of Eq. (\ref{eq:2.2}) with $\beta \rightarrow 
\beta_0$. Since $\delta \approx 0.27 << 1 $, we can expand the J-integrals
in powers of $\delta$ and retain terms up to linear in $\delta$. Thus,
for each mechanism (vector, scalar, spin) we obtain two integrals 
associated with $l=0$ and $l=1.$ The most important $K^l$ integrals are 
shown in Fig. 1(b). For some others see ref.~\cite{KVprd} By exploiting the 
above expansion of K-integrals, the counting rate can be written in the
form

\beq
\Big<\frac{dN}{dt}\Big>=\Big<\frac{dN}{dt}\Big>_{0}(1 + h \, cos\alpha)
\label{eq:3.3}
\eeq

\noindent
where $\alpha$ is the phase of the earth's orbital motion and 
$\big<dN/dt\big>_0$ is the rate obtained from the $l=0$ multipole. 
$h$ represent the amplitude of the oscillation, i.e. the ratio of the 
component of the multipole $l=1$ to that of the multipole $l=0$. 

\section{Results and discussion }
The three basic ingredients of our calculation were 
the input SUSY parameters, a quark model for the nucleon and the structure
of the nuclei involved. The input SUSY parameters used  
have been calculated in a phenomenologically allowed 
parameter space (cases \#1, \#2, \#3) as explained in ref.~\cite{KVprd}

In  Tables 1, we compare the spin matrix elements at $q=0$ for the 
most popular targets considered for LSP detection $^{207}Pb$, 
$^{73}Ge$ and $^{29}Si$. We see that, the spin matrix elements 
$\Omega^2$ of $^{208}Pb$ are even a factor of three smaller than those 
for $^{73}Ge$ obtained in ref.~\cite{Ress}

The spin contribution, arising from the axial current,
was computed in the case of $^{207}Pb$ system (see Tables 2, 3 and 4). 
For the isovector axial coupling the transition from the quark to
the nucleon level is trivial (a factor of $g_A=1.25)$. For the
isoscalar axial current we considered two possibilities 
depending on the portion of the nucleon spin which is attributed to the
quarks, indicated by EMC and NQM.~\cite{KVprd} The ground state wave 
function of $^{208}Pb$ was obtained by diagonalizing the nuclear 
Hamiltonian in a 2h-1p space which is standard for this doubly magic nucleus.   

\begin{table}[t]
\caption{Comparison of the static spin matrix elements for three
typical nuclei.\label{tab:1}}
\vspace{0.4cm}
\begin{center}
\footnotesize
\begin{tabular}{|l|rrr|}
\hline
Component & $^{207}Pb_{{1/2}^-}$ & $^{73}Ge_{{9/2}^+}$ 
 & $^{29}Si_{{1/2}^+}$ \\
\hline
$\Omega^2_1(0)$ & 0.231 & 1.005 & 0.204 \\
$\Omega_1(0) \Omega_0(0)$ & -0.266 & -1.078 & -0.202 \\
$\Omega^2_0(0)$ & 0.305 & 1.157 & 0.201 \\
\hline
\end{tabular}
\end{center}
\end{table}

\begin{table}[t]
\caption{The spin contribution in the $LSP-^{207}Pb$ scattering 
for two cases: EMC data and NQM Model. The LSP mass is
$m_1=126, \,\,27, \,\,102 \,\,GeV$ for $\#1, \#2, \#3$ 
respectively. \label{tab:2}}
\vspace{0.4cm}
\begin{center}
\footnotesize
\begin{tabular}{|l|ll|lc|}
\hline
& \multicolumn{2}{|c}{\hspace{1.2cm}EMC \hspace{.2cm} DATA} \hspace{.8cm} &
 \multicolumn{2}{c|}{\hspace{1.2cm}NQM \hspace{.2cm} MODEL} \\ 
\hline
Solution & \hspace{.2cm}$< dN/dt >_0 $ $(y^{-1} Kg^{-1})$
  \hspace{.2cm} & $ h $ & 
$\hspace{.2cm} < dN/dt >_0 $ $ (y^{-1} Kg^{-1})$ & $ h $ \\ 
\hline
$\#1  $  &$0.285\times 10^{-2}$& 0.014 &$0.137\times 10^{-2}$& 0.015  \\
$\#2  $  & 0.041               & 0.046 &$0.384\times 10^{-2}$& 0.056  \\
$\#3  $  & 0.012               & 0.016 &$0.764\times 10^{-2}$& 0.017  \\
\hline
\end{tabular}
\end{center}
\end{table}

For the coherent part (scalar and vector) we used realistic nuclear
form factors. In ref.~\cite{KVprd} we studied three nuclei, representaves
of the light, medium and heavy nuclear isotopes ($Ca$, $Ge$ and $Pb$)
for three different quark models. In table 4 we present the results for $Pb$
obtained for these models denoted by A (only quarks u and d in the nucleon)
and B and C (heavy quarks included) for the LSP
masses shown in table 3 (cases \#4-\#9).
We see that the results vary substantially and are sensitive to 
the presence of quarks other than u and d into the nucleon.

The total cross section is almost the same for LSP masses around $100\,GeV$. 
We obtained results in the context of the quark models NQM, EMC,
for SUSY models \#1-\#3~\cite{Kane} (Tables 2 and 3) and SUSY models 
\#4-\#9~\cite{Casta} (Tables 4 ).

\begin{table}[t]
\caption{Ratio of spin contribution ($^{207}Pb/^{73}Ge$) at the relevant
momentum transfer with the kinematical factor $1/(1+\eta)^2, \,\, 
\eta=m_1/A m_p.$
\label{tab:3}}
\vspace{0.4cm}
\begin{center}
\footnotesize
\begin{tabular}{|c|lllllllll|}
\hline
 Solution & $\#$1 & $\#$2 & $\#$3 & $\#$4 & $\#$5 & $\#$6 & $\#$7 & $\#$8 
 & $\#$9 \\
\hline
$m_1 \,(GeV)$ & 126 & 27 & 102 & 80 & 124 & 58 & 34 & 35 & 50 \\
\hline
NQM & 0.834 & 0.335 & 0.589 & 0.394 & 0.537 & 0.365 & 0.346 & 0.337 & 0.417 \\
EMC & 0.645 & 0.345 & 0.602 & 0.499 & 0.602 & 0.263 & 0.341 & 0.383 & 0.479 \\
\hline
\end{tabular}
\end{center}
\end{table}

\begin{table}[t]
\caption{Event rates for $^{207}Pb$. The LSP mass
is $m_1= 80,$ 124, 58, 34, 35, 50 $GeV$ for the cases $\#4-\#9$ 
respectively. Cases $\#8, \#9$ are no-scale models. The values of 
$<dN/dt>_0$ for Model A and the Vector part must be multiplied by
$\times 10^{-2}$.
\label{tab:4}}
\vspace{0.4cm}
\begin{center}
\footnotesize
\begin{tabular}{|l|crrr|lr|lll|}
\hline
  & \multicolumn{4}{|c|}{Scalar$\,$ Part} & 
    \multicolumn{2}{c|}{Vector $\,$ Part} &
    \multicolumn{3}{c|}{Spin   $\,$ Part} \\
\hline 
& \multicolumn{3}{c}{$\big<\frac{dN}{dt}\big>_0$}& $h$ & 
$\big<\frac{dN}{dt}\big>_0$ & $h$ & \multicolumn{2}{c}
{$\big<\frac{dN}{dt}\big>_0$} & $h$ \\ 
\hline
& A & B & C & & & & EMC & NQM & \\ 
\hline
$\#4 $& 0.03 & 22.9& 8.5 & 0.003& 0.04 & 0.054
 & $0.80\, 10^{-3}$& $0.16\, 10^{-2}$& 0.015 \\
$\#5 $& 0.46 & 1.8& 1.4 & -0.003& 0.03 & 0.053
 & $0.37\, 10^{-3}$& $0.91\, 10^{-3}$& 0.014 \\
$\#6 $& 0.16 & 5.7& 4.8 & 0.007& 0.11 & 0.057
 & $0.44\, 10^{-3}$ & $0.11\, 10^{-2}$& 0.033 \\
$\#7 $&  4.30 & 110.0& 135.0 & 0.020& 0.94 & 0.065
 & 0.67 & 0.87 & 0.055 \\
$\#8 $& 2.90 & 73.1& 79.8 & 0.020& 0.40 & 0.065
 & 0.22 & 0.35 & 0.055 \\
$\#9 $&  2.90 & 1.6& 1.7 & 0.009& 0.95 & 0.059
 & 0.29 & 0.37 & 0.035 \\
\hline
\end{tabular}
\end{center}
\end{table}

\section{Conclusions }
In the present study we found that for heavy LSP and heavy nuclei
the results are sensitive to the momentum transfer as well as to the
LSP mass and other SUSY parameters. From the Tables 3 and 4 we see that,
the results are also sensitive to the quark structure of the nucleon
in the following sense: i) The coherent scalar (associated with Higgs 
exchange) for model A (u and d quarks only) is comparable to the 
velocity suppressed vector coherent contribution.
Both are at present undetectable. ii) For models B and C (heavy
quarks in the nucleon) the coherent scalar contribution is
dominant. Detectable rates $<dN/dt>_0 \ge 100\, \, y^{-1}Kg^{-1}$ 
are possible in a number of models with light LSP.

The spin contribution is sensitive to the nuclear structure.
It is undetectable if the LSP is primarily a gaugino. 

The folding of the event rate with the velocity distribution provides 
the modulation effect $h$. In all cases it is small, less than $\pm 5\%$.


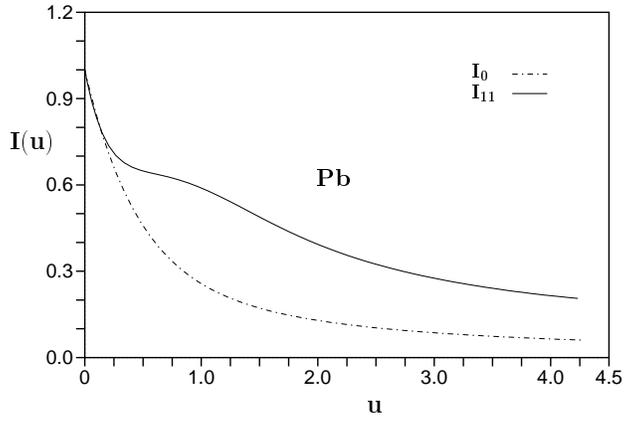
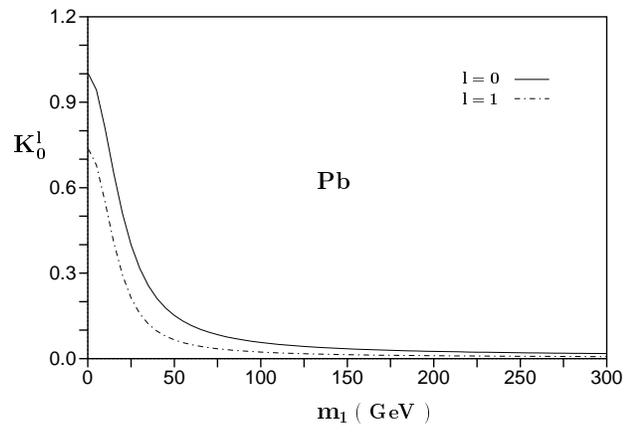
\end{document}